\documentclass[aps,prb,twocolumn,superscriptaddress,floatfix,showpacs]{revtex4}
\usepackage{graphicx}
\usepackage{bm}
\usepackage{color}
\usepackage{longtable}
\usepackage{amssymb}
\usepackage{amsmath}
\bibstyle{apsrev.bib}

\begin{document}
\title{Logarithmic delocalization of end spins in the $S=3/2$ \\antiferromagnetic Heisenberg chain}
\author{G\'abor F\'ath}
\affiliation{Research Institute for Solid State Physics and Optics,
H-1525 Budapest, P.O.Box 49, Hungary}
\author{\"Ors Legeza}
\affiliation{Research Institute for Solid State Physics and Optics,
H-1525 Budapest, P.O.Box 49, Hungary}
\author{P\'eter Lajk\'o}
\affiliation{Department of Physics, Kuwait University,
P. O. Box 5969, Safat, Kuwait 13060}
\author{Ferenc Igl\'oi}
\affiliation{Research Institute for Solid State Physics and Optics,
H-1525 Budapest, P.O.Box 49, Hungary}
\affiliation{Institute of Theoretical Physics, Szeged University,
H-6720 Szeged, Hungary}

\date{\today}

\begin{abstract}
Using the DMRG method we calculate the surface spin correlation function, $C_L(l)=\langle S_l^z
S_{L+1-l}^z \rangle$, in the spin $S=3/2$ antiferromagnetic Heisenberg
chain. For comparison we also investigate the $S=1/2$ chain with $S=1$ impurity end spins and the  $S=1$ chain. In the half-integer spin models the end-to-end correlations are found to decay to zero logarithmically, $C_L(1)\sim (\log L)^{-2d}$, with $d=0.13(2)$. We find no surface order, in clear contrast with the behavior of the $S=1$ chain, where exponentially localized end spins induce finite surface correlations. The lack of surface order implies that end spins do not exist in the strict sense. However, the system possesses a logarithmically weakly delocalizing boundary excitation, which, for any chain lengths attainable numerically or even experimentally, creates the illusion of an end spin. This mode is responsible for the first gap, which vanishes asymptotically as
$\Delta_1 \approx  (\pi v_S d)/(L\ln L)$, where $v_S$ is the sound velocity and $d$ is
the logarithmic decay exponent. For the half-integer spin models our results on the surface correlations and on the first gap support universality. Those for the second gap are less conclusive, due to strong higher-order corrections.
\end{abstract}


\maketitle

\newcommand{\bc}{\begin{center}}
\newcommand{\ec}{\end{center}}
\newcommand{\be}{\begin{equation}}
\newcommand{\ee}{\end{equation}}
\newcommand{\ba}{\begin{array}}
\newcommand{\ea}{\end{array}}
\newcommand{\beqn}{\begin{eqnarray}}
\newcommand{\eeqn}{\end{eqnarray}}

\section{Introduction}

One of the most studied models of interacting quantum systems is the
antiferromagnetic Heisenberg chain. It is defined by the
Hamiltonian
\begin{equation}
{\cal H}_S = \sum_{i=1}^{L-1} \vec{S}_i \cdot \vec{S}_{i+1}\;,
\label{hamilton}
\end{equation}
where $\vec{S}_i$ is the spin-$S$ operator at
site $i$.  It is known since Haldane's seminal work\cite{Hald81} that
the low-energy properties of the model for half-integer and integer
values of the spin are different. Half-integer spin chains are gapless,
and the bulk correlations are quasi-long-ranged. On the other hand,
integer spin chains have a finite gap, there is a hidden topological
order,\cite{denN89} and the bulk correlations are short-ranged,
involving a finite correlation length $\xi$.

The universality class of half-integer spin chains was debated for
some time, by now it is generally accepted that all are in the same
bulk universality class, independent of the value of the
spin.\cite{s32,AfHa,AGST} A thorough numerical investigation for the $S=3/2$
chain\cite{Hall96} verified convincingly that the bulk spin-spin correlation function
$\langle S^{z}_0 S^{z}_r \rangle\sim r^{-\eta}$ decays
with the same \emph{bulk exponent} $\eta=1$ as that of the $S=1/2$ model.
It is a priori not clear, however, to what extent this universality can hold
for the surface behavior.

In an open chain with $L$ sites it is of
interest to measure the two-point function,
\be
C_L(l)=\langle S^{z}_l S^{z}_{L+1-l} \rangle\;,
\label{cij}
\ee
where the two spins are positioned symmetrically to the middle of the chain. When $1\ll l<L/2$ the two spins are far from the boundary and the scaling behavior is determined by the bulk exponents. For instance, $C_L(\alpha L)$ with $0\ll \alpha< 1/2$ scales as $C_L(\alpha L)\sim L^{-\eta}$.

On the other hand, for $l\ll L$ fixed, the two spins are close to the boundary, and $C_L(l)$ probes the surface critical behavior. In particular, for $l=1$ we obtain the \emph{end-to-end correlation function} $C_L(1)$.
%
%
For the $S=1/2$ chain, as obtained by analytical methods, the end-to-end
correlation is also quasi-long-ranged, but the asymptotic decay,
$C_L(1) \sim L^{-\eta_s}$, involves the so-called
\emph{surface exponent} $\eta_s=2$,\cite{ABGR} which differs from the
bulk exponent $\eta$.

For higher $S$ there may be an even more drastic difference between bulk and surface correlations; the surface correlations may become long-ranged. This phenomenon is best understood for the $S=1$ chain,
in which there is a well-established \emph{surface order}, $C(1)=\lim _{L \to \infty} C_L(1)>0$. In general the order decays exponentially with the distance from the surface,
\be
C(l) = \lim_{L \to \infty} C_L(l) \sim \exp(-2 l/\xi_s)\;.  
\label{C_l_1}
\ee
where $\xi_s$ is the surface correlation length, which is equal to the bulk
correlation length $\xi$ of the $S=1$ chain.\cite{white_huse,sorensenaffleck94} (Note, however, the factor $2$ in the exponent, which is present due to the fact that a unit change in $l$ takes the two spins two sites closer to each other.)
The leading finite-size correction to the correlation function is also exponential,
\be
C_L(l)-C(l) \sim \exp(-L/\xi_f)\;.   
\label{C_1_L}
\ee
Here again the length $\xi_f$ characterizing the finite-size corrections is
expected to be equal to $\xi$, thus there is only one length scale in
the system. Note that we have neglected possible algebraic prefactors in
Eqs.\ (\ref{C_l_1}) and (\ref{C_1_L}).

The origin of surface long-range order in the $S=1$ chain can be
understood in the valence-bond-solid (VBS) framework.\cite{AKLT}
Representing each $S=1$ spin by two symmetrically coupled spin-$1/2$
degrees of freedom, the ground state can be written as the product of
singlet bonds connecting neighboring spins, except for two remaining
localized spin-$1/2$s at the ends. These localized end spins are
responsible for the behavior of surface spin correlations. The VBS construction is exact for the $S=1$ Affleck-Kenedi-Lieb-Tasaki (AKLT) model,
\begin{equation}
{\cal H}_{\rm AKLT} = \sum_{i=1}^{L-1} \vec{S}_i \cdot \vec{S}_{i+1} +\frac{1}{3}\left(\vec{S}_i \cdot \vec{S}_{i+1}\right)^2\;.
\label{aklt_hamilton}
\end{equation}
However, numerical studies have confirmed the above scenario for the pure $S=1$ chain, ${\cal H}_1$, too.\cite{sorensenaffleck94}

In order to understand the behavior of open spin chains with $S>1$, Tai-Kai Ng proposed\cite{ng94} a theory based on an extension of the VBS picture. According to this, each spin-$S$ variable can be replaced by $2S$ symmetrically coupled spin-$1/2$s, among which singlet bonds are
formed. For integer $S$ all bulk spins are involved in singlet
formation, except for those at the chain ends, which form localized edge spins
with spin $S/2$. Surface spin correlations are then expected to have
the same characteristics as for $S=1$, summarized in Eqs.\
(\ref{C_l_1}) and (\ref{C_1_L}).  The ground state becomes $(S+1)^2$ times degenerate in the thermodynamic limit. For finite chains this degeneracy is lifted by an exponentially small gap, which is due to the exponentially small effective interaction between the end spins. Above the degenerate ground states the spectrum is gapped (Haldane gap), and
the finite-size corrections to these levels are algebraic (note the boundary condition).\cite{sorensenaffleck94} DMRG-studies of the $S=2$ spin chain have shown evidence
in favor of the above scenario, although the correlation length is
much longer than for $S=1$.\cite{ng95,qin97,wang99}

For half-integer spins, this scenario implies that there is one
spin-$1/2$ at each bulk site which is not involved in singlet
formation. Moreover, there remain spin-$(S+1/2)/2$ degrees
of freedom at the chain ends. Consequently, as Ng argues,\cite{ng94} the effective low-energy
model is a spin-$1/2$ chain (a Luttinger liquid) perturbed by $(S+1/2)/2$ impurity spins coupled antiferromagnetically at the ends:
\begin{equation}\label{equiv1}
    {\cal H}_{S}\sim {\cal H}_{1/2}^{(S+1/2)/2}(\lambda>0), \quad \mbox{$S=$half-integer},
\end{equation}
where
\begin{equation}
{\cal H}_{1/2}^{\sigma}(\lambda) =
                        \sum_{i=2}^{L-2} \vec{S}_i \cdot \vec{S}_{i+1} +
                       \lambda\left( \vec{\sigma}_1 \cdot \vec{S}_{2} + \vec{S}_{L-1} \cdot \vec{\sigma}_L \right),
\label{hamilton2}
\end{equation}
with $\vec{S}$ ($\vec{\sigma}$) being spin-1/2 (spin-$\sigma$) operators, and $\lambda>0$ is some unknown, but positive, impurity coupling.

The ${\cal H}_{1/2}^{\sigma}$ quantum impurity problem was analyzed earlier by
Eggert and Affleck.\cite{Eggert_Affleck92} They argued that the surface spin is partially screened by the Luttinger liquid, leading to an effective end spin of value $\sigma_{end}=\sigma-1/2$. This remaining spin is coupled ferromagnetically to the bulk, thus
\begin{equation}\label{equiv2}
    {\cal H}_{1/2}^{\sigma}(\lambda>0)\sim {\cal H}_{1/2}^{\sigma-1/2}(-g<0),
\end{equation}
as far as the low-energy behavior is considered.
However, the coupling $g=g(L)>0$ is marginally irrelevant, satisfying a Kondo-type RG equation
\begin{equation}\label{RG}
    \frac{dg}{d\ln L} \sim -g^2,
\end{equation}
whose solution is
\begin{equation}\label{g_scales}
    g(L)\sim \frac{1}{\ln(BL)}.
\end{equation}
Here $B$ is an appropriate inverse length depending on the bare coupling. Note that $B$ can be rather different from unity, thus its effect cannot be neglected for moderately long chains.

Equation (\ref{g_scales}) implies that end spins become asymptotically free, $g(L)\to 0$ for $L\to \infty$. 
In the thermodynamic limit the spectrum is that of an $S=1/2$ chain with $L-2$ sites and two decoupled spin-$\sigma_{end}$ spins. This gives rise to an $(2\sigma_{end}+1)^2$-fold quasi-degeneracy of the ground state. For finite length the quasi-degenerate states are separated from the ground state by an energy
\be
    \Delta_{\rm deg}(L) \sim \frac{g(L)}{L}\sim \frac{1}{L\ln(BL)}.
    \label{e1_log}
\ee
%
The other higher-lying excitations in the spectrum are predicted to be delocalized excitations of the Luttinger liquid.

Some of these field-theoretical predictions have been checked
numerically for half-integer spin Heisenberg chains.\cite{ng95,qin97,lou_prb00,lou_prb02,legeza_prb99} In particular, indications for the predicted number of edge states were found, and the corresponding excitation energies were shown to vanish faster than $1/L$. Higher-lying states were demonstrated to scale as $1/L$. However, several aspects of the system have remained unclear. The accurate form of the excitation energy of the edge state lacks an accurate analysis. It is not clear whether the logarithmic term predicted by the quantum
impurity calculation is indeed consistent with the numerical results.
Also, in order to test universality the structure of the higher lying excitations requires additional analysis.

Another open question involves the surface correlation function, and in particular the surface long-range order. Indeed, the VBS analysis is based on the assumption that the bulk chain possesses a short-range RVB "backbone" which remains inert for the low-energy behavior. Its only effect is to provide localized edge spins. This picture would imply that this backbone mediates correlation between the two ends of the chain in much the same way as it does for the $S=1$ Heisenberg chain. However, the presence of the extra spin-1/2 component in the bulk may have an important effect, and it is a priori unclear whether this effect is weak enough not to destroy the surface order. In Ng's theory the edge spins are localized to the ends, and their effective coupling scales to zero faster than $1/L$. When the chain length is even, the lowest-lying state is a singlet, in which both the bulk of the chain and the two end spins form singlets independently. Thus there is a clear anticorrelation between end spins. Localization of the end spins means that they contribute a finite amount to the first and last spins of the chain, thus this anticorrelation assures finite surface order, $C(l)>0$.

Our aim in this paper is to elaborate on these questions in the case of the $S=3/2$ antiferromagnetic Heisenberg chain. We study the spin-3/2 chain, ${\cal H}_{3/2}$, using the DMRG method,
and calculate low-lying excitation energies and surface correlation function. For comparison with the VBS theory, we
also consider the $S=1$ chains, ${\cal H}_{AKLT}$ and ${\cal H}_{1}$, and the effective model, ${\cal H}_{1/2}^{1}(\lambda>0)$, which is predicted by Ng's theory to belong to the same universality class as the $S=3/2$ chain.

The structure of the paper is the following. In Sec.\ \ref{sec:num} we discuss the high precision DMRG method used in the calculations. The numerically calculated excitation energies and the surface correlation function are analyzed in Sec.\ \ref{sec:gap}
and in Sec.\ \ref{sec:corr}, respectively. A discussion of the results is presented in Sec.\ \ref{sec:disc}. There is an Appendix attached to the paper, where we list the exact surface critical properties of an Ising quantum chain with an extended defect, which serves as an analogy with rigorous results.

\section{Numerical method}
\label{sec:num}

In the calculations we used open chains of various finite lengths, $L=12,24,36,\dots,216$, and
calculated energy gaps and the surface correlation function using
the DMRG method (for recent reviews see Ref.\ \onlinecite{DMRGbook}). We
performed the DMRG calculations by using both the standard technique,
i.e., keeping the number of block states fixed, and the dynamic block
state selection (DBSS) approach.\cite{legeza02,legeza03} All
eigenstates of the model were targeted independently using two or
three DMRG sweeps.

In the standard procedure $500-800$ block states
were used. It was found that for the largest systems ($L=192$ for the
$S=3/2$ chain and $L = 216$ for the $S=1/2$ chain with impurities), the truncation
error varied in the range $10^{-7}-10^{-8}$.  The DBSS
approach\cite{legeza02} allows for a more rigorous control of
numerical accuracy, and we set the threshold value of the quantum
information loss $\chi$ to $10^{-8}$. The minimum number of block
states was set to $256$.  The entropy sum rule was checked for all
finite chain lengths for each DMRG sweeps, and it was found that the
sum rule was satisfied after the second sweep already. The maximum
number of block states varied in the range $600-1400$. In order to have high precision results, we set the residual error of the Davidson diagonalization algorithm to $10^{-9}$.

\section{Excitation energies}
\label{sec:gap}

The spectrum of the $S=1$ chain with open boundaries is well known in
the literature.\cite{kennedy} The first gap (singlet-triplet gap), which is related to the
localized surface excitation, is exponentially small in $L$. In contrast with this, the second gap (singlet-quintuplet gap), which describes a bulk excitation, is finite and this defines the well-known Haldane gap.

For $S=3/2$, numerical results on the spectrum of open
chains were presented in Ref.\ \onlinecite{ng95}, in which the authors
aimed to make a comparison with the field-theoretical prediction of
Ng.\cite{ng94} Indeed the numerical results of Ref.\ \onlinecite{ng95}
are compatible with the prediction that the first gap vanishes faster
than $1/L$. However, as their quantitative analysis of the finite-size
dependence of the first gap was rather sketchy, we are going to repeat this
type of analysis here. Furthermore, we make a comparative study of the
spectrum of the effective model ${\cal H}_{1/2}^1(\lambda)$, which arises from the VBS picture. For $\lambda$, which is an a priori unknown positive bare coupling, we test three different values, $\lambda=0.5,1.0$ and $4.0$. We expect that the actual value does not matter due to universality.

In both half-integer spin models the remaining effective end spins are $\sigma_{end}=1/2$, thus the quasi-degeneracy involves the singlet ground state and a triplet state. Thus the first gap, defined by the lowest state in the $S^z_{tot}=1$ sector, is expected to take the form of Eq.\ (\ref{e1_log}).
The higher-lying excitations are delocalized excitations. They scale as $1/L$ and form conformal towers. In the following we analyze the gaps $\Delta_n$, $n=2,3,\dots$, defined by the lowest states in the sectors $S^z_{tot}=n$.

Conformal invariance dictates that the gap $\Delta_n$, associated with the scaling field $\phi_n$,  satisfies
\begin{equation}\label{cft}
    \Delta_n=\frac{\pi v_{S}}{L} \left[ x_{n}+\frac{d_n}{\ln(BL)} +
    {\cal O}\left(\frac{\ln\ln L}{\ln^2 L}\right)   \right] \;,
\end{equation}
where $v_S$ is the non-universal sound velocity and $x_n$ is the scaling dimension of $\phi_n$, defined as $\langle \phi_n(0)\phi_n(r)\sim r^{-2x_n}$. The sound velocity is known exactly for the $S=1/2$ chain: $v_{1/2}=\pi/2$.\cite{vs} For $S=3/2$ the best numerical estimate is $v_{3/2}=3.87$.\cite{Hall96}

The logarithmic correction appears due to the presence of marginal operators $\phi_{\rm marg}$.\cite{cardy_jpa86} The coefficient $d_n$ is universal, determined unambiguously by the operator product expansion (OPE) rules of the underlying field theory.\cite{cardy_jpa86,cardy_book} Assuming that the OPE reads schematically as
\begin{eqnarray}
    \phi_{\rm marg}\cdot\phi_{\rm marg} &=& b\, \phi_{\rm marg} +\dots \nonumber\\
    \phi_n\cdot\phi_n &=& b_n\, \phi_{\rm marg} +\dots,
\end{eqnarray}
the coefficient of the log correction turns out to be\cite{cardy_jpa86}
\begin{equation}\label{dn}
    d_n = \frac{2b_n}{b}.
\end{equation}
Other, more irrelevant operators may also give corrections but these decay faster than the leading order logarithmic correction. In the present case there are, in fact, two marginal operators in the theory: the boundary perturbation represented by the effective edge spin discussed above, and the standard bulk marginal operator present in all half-integer spin chains.\cite{AGST} In the following we assume that even in this case Eq.\ (\ref{cft}) remains valid, with $d_n$ representing an effective value, arising due to the combined effect of the two marginal operators.


Note that the expected form of the gaps in Eqs.\ (\ref{e1_log}) and (\ref{cft}) are formally the same. We can think about the first gap as satisfying Eq.\ (\ref{cft}) with scaling dimension $x_1=0$ and an appropriate constant $d_1$.

Neither for ${\cal H}_{3/2}$ nor for the ${\cal H}_{1/2}^1(\lambda)$ model are we aware of any rigorous analytical predications for $x_n$ and $d_n$. Nonetheless, if the two models are in the same (surface) universality class, the full operator content should be identical. A possibility, suggested implicitly in Ref.\ \onlinecite{ng95} is that the critical exponents associated with the higher-lying excitations in the spectrum are the same as the exponents in the pure $S=1/2$ model with open boundary condition, ${\cal H}_{1/2}$. If this were true, the edge spins would only effect the lowest gap $\Delta_1$ but not any higher-lying excitations. For the ${\cal H}_{1/2}$ chain the operator content is rigorously known:\cite{ABGR,affleck_jpa99}
\begin{eqnarray}\label{opcontent}
    x_n &=& n^2 \nonumber\\
    d_n &=& -n(n+1)/2.
\end{eqnarray}
However, numerical results of Ref.\ \onlinecite{ng95}, when adjusted correctly with the sound velocity $v_{3/2}$, do not support convincingly this scenario for the $S=3/2$ chain. It may well be that the end perturbation, which has similar effect to a modified boundary condition, substantially alters the surface operator content. This phenomenon is in fact possible: ${\cal H}_{1/2}$ (open) and ${\cal H}_{1/2}+\vec{S}_L \cdot \vec{S}_1$ (periodic) have different operator content.\cite{ABGR,affleck_jpa99}
Nevertheless, we can use the values in Eq.\ (\ref{opcontent}) as reference values for comparison in the sequel.


\subsection{Analysis of the first gap}

\begin{figure}[tbp]
\centerline{\includegraphics[scale=.33]{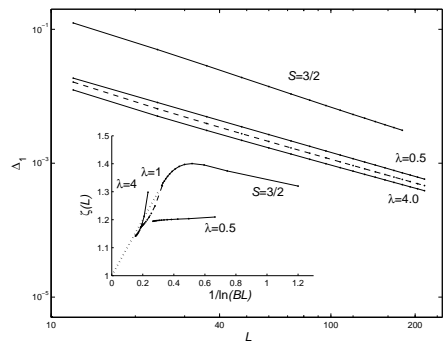}} 
\caption{Log-log plot of the first gap $\Delta_1$ vs. $L$; The inset shows the associated local exponent $\zeta(L)$ vs. $1/\ln (BL)$ with $B$ chosen appropriately for each data set.}
\label{Fig3}
\end{figure}

The size dependence of the first gap is shown in Fig.\ \ref{Fig3}
both for ${\cal H}_{3/2}$ and for the effective model ${\cal H}_{1/2}^1(\lambda)$ with three
different values of the impurity coupling, $\lambda=0.5,1.0,4.0$. The data
seemingly fit onto a straight line in the log-log plot, with an
average slope larger than 1. To obtain a more quantitative
description, we have calculated the \emph{local slopes} from two point
fits.  This corresponds to the effective gap exponent $\zeta(L)$,
defined as $\Delta_1(L) \sim L^{-\zeta(L)}$, and the true gap exponent
is obtained as $\zeta=\lim_{L \to \infty} \zeta(L)$.

For the $S=3/2$ chain, the effective exponent first
increases with $L$, then passing through a maximum around $L=60$, it starts to
decrease rapidly. The same data for the impurity models are
monotonically decreasing with $L$. To analyze the
size dependence of the effective gap exponents we use the
field-theoretical result in Eq.\ (\ref{e1_log}), which predicts
logarithmic corrections. Therefore the effective exponents are
presented in the inset of Fig.\ \ref{Fig3}(a) as a function of
$1/\ln(BL)$ with the optimal $B$ chosen independently for each data
set (see later).  For almost all curves the behavior seems
to be nicely compatible with the expected result $\zeta= 1$ with logarithmic corrections, except for $\lambda=0.5$ where the asymptotic regime has probably not yet been reached.

\begin{figure}[tbp]
\centerline{\includegraphics[scale=.42]{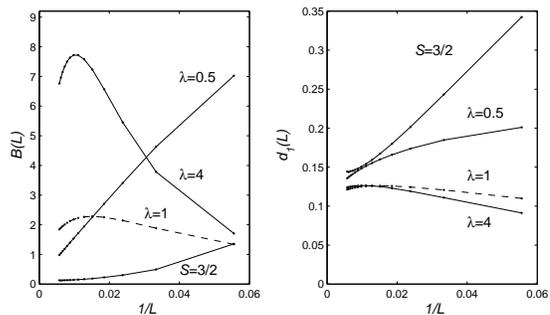}} 
\caption{Length-dependent effective parameters, $B(L)$ and $d_1(L)$ determined from two-point fits. Dotted line segments indicate uncertainty.}
\label{Fig_Bd1}
\end{figure}

Assuming that the first gap indeed scales according to Eq.\ (\ref{cft}) with $x_1=0$, the two remaining parameters $B$ and $d_1$ can be determined from the numerical data. The consistency of these parameters with other results can serve as a justification for the fitting ansatz. In Fig.\ \ref{Fig_Bd1} we present the effective, size-dependent parameters $B(L)$ and $d_1(L)$, determined from two-point fits. The parameter $B$ which depends on the bare couplings of the marginal operators is not universal. Indeed we see that it varies considerably for the four models we consider. Our best estimates for $B=\lim_{L\to\infty} B(L)$ are as follows:
\begin{equation}\label{Bs}
   B=\left\{ \begin{array}{ll}
       0.11\pm 0.02 & \mbox{for ${\cal H}_{3/2}$} \\
       0.25\pm 0.1 & \mbox{for ${\cal H}_{1/2}^1(\lambda=0.5)$} \\
       1.2\pm 1 & \mbox{for ${\cal H}_{1/2}^1(\lambda=1)$} \\
       4\pm 2 & \mbox{for ${\cal H}_{1/2}^1(\lambda=4)$} \\
    \end{array} \right.
\end{equation}
Note, however, that $B$ is an extremely sensitive parameter since it is in the argument of a logarithm. Even the $10^{-10}$ relative precision of the gap values may not be enough for a reliable extrapolation. Indeed, we cannot exclude that the sharp downturn of $B(L)$ for the impurity models above $L\approx 72$ may indicate a systematic error stemming from the loss of numerical precision, and not a real physical effect. Thus the determination of $B$ for these models bears a rather large error.

Even though the error of $B$ is considerable, many other quantities do not inherit this large error bar due to their logarithmic dependence on $B$. This is clearly seen for $d_1$ in Fig.\ \ref{Fig_Bd1}. Our best estimate for ${\cal H}_{3/2}$ is $d_1=0.14\pm 0.02$, and for ${\cal H}_{1/2}^1$ $d_1=0.12\pm 0.02$. The deviation is small, and a common value at $d_1=0.13\pm 0.02$ is consistent with our error estimates. This supports the conjecture that these models belong to the same universality class. Note that $d_1$ is positive, whereas for the pure $S=1/2$ chain, ${\cal H}_{1/2}$, $d_n$ in Eq.\ (\ref{opcontent}) is negative for all $n$.

\begin{figure}[tbp]
\centerline{\includegraphics[scale=.32]{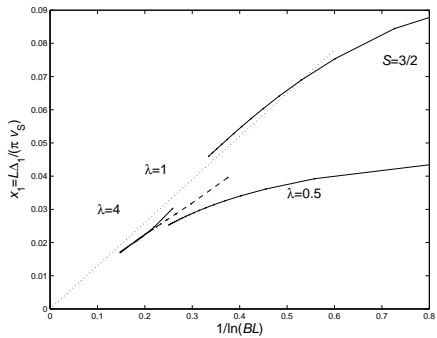}} 
\caption{Scaled first gap $L\Delta_1$ as a function of $1/\ln(BL)$, with $B$ chosen individually for each curve. The common slope (dotted line) determines the universal value $d_1=0.13\pm 0.02$.}
\label{Fig_scaledgap1}
\end{figure}

Using the values of $B$ determined in Eq.\ (\ref{Bs}), the first scaled gap is convincingly linear as a function of $1/\ln(BL)$ as shown in Fig.\ \ref{Fig_scaledgap1}. The common slope is given by $d_1$ within the error bar.

The corresponding value $d_1$ for the $S=3/2$ chain can also be calculated from the
results of Lou \emph{et al.}\cite{lou_prb00} on the coefficient of the
log correction. Their data yields $d_1=0.32$, however, they used a very rough fitting with $B=1$,
which is very far from our estimated value.
As for $B$ itself, our value can be
compared to the one obtained in Ref.\ \onlinecite{Hall96}, namely
$B=0.6\pm 0.1$. The two values differ by a factor of five. Notice, however, that Ref.\ \onlinecite{Hall96} used periodic boundary condition, and thus measured the logarithmic corrections due to the bulk marginal operator only, whereas our $B$ is expected to represent the combined effect of both marginal operators. Also $B(L)$ is monotonically decreasing for $S=3/2$, thus determining it from shorter chains such as $L=60$ used in Ref.\ \onlinecite{Hall96}, may yield a higher effective value.

\subsection{Analysis of the second gap}

\begin{figure}[tbp]
\centerline{\includegraphics[scale=.32]{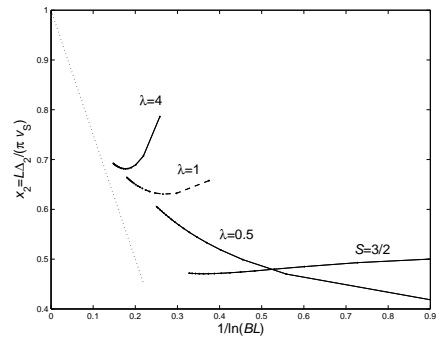}} 
\caption{Scaling dimension $x_2$ associated with the scaled second gap
$\Delta_2 L$ as a function of $1/\ln BL$. The dotted line indicates the conjectured scaling.}
\label{Fig4}
\end{figure}

Our numerical data indicate both for the $S=3/2$ chain and for the
impurity models that the second gap $\Delta_2$ scales as $1/L$, and the
associated dimension $x_2$ is finite. The scaled second gap $L\Delta_2/(\pi v_S)$ is shown in Fig.\
\ref{Fig4} as a function of $1/\ln(BL)$. For $B$ we have used the
values determined from the first gap. It is clear that the finite size
corrections are very strong and the asymptotic regimes have not yet been reached.

The curves for ${\cal H}_{1/2}^1(\lambda)$ go through a local minimum, then increase rapidly as $L$ increases. A limiting value around $x_2\approx 1$ seems plausible, but the approximation error is large, and in fact nothing above 0.8 can be firmly excluded. As mentioned above the scaling dimension $x_2=1$ would be expected if the delocalized excitations behaved as those of the pure ${\cal H}_{1/2}$ model for which the lowest scaling dimension given by Eq.\ (\ref{opcontent}) is 1.

The coefficient of the leading logarithmic correction $d_2$ may again be universal in accordance with Eq.\ (\ref{cft}). The most probable value, consistent with our data is $d_2=-2.5\pm 1$.
This has different sign than $d_1$. It also differs considerably from $d_1=-1$ characterizing the lowest gap in ${\cal H}_{1/2}$. Nevertheless, we should warn the Reader that the estimation of $x_2$ and $d_2$ depends very much on the precision of $B$. Changing $B$ practically acts as a horizontal shift in the Figure which can modify substantially the extrapolation results.

For the $S=3/2$ chain, as seen in Fig. \ref{Fig4},
the finite-size dependence is very small. The scaled gap initially
decreases in $L$, then starts to increase very slightly for larger $L$. If
this were the thermodynamic limit, we would conclude that $x_2\approx
0.47\pm 0.01$. This would contradict universality with the impurity model.
However, having seen the nonmonotonic behavior for the impurity model,
there is some doubt for this conclusion. We have to leave open the
possibility that the local minimum of $x_2$ may be followed by a
strong upturn for larger values of $L$. We estimate that the crossover length for this upturn should be, in worst case, around $1/\ln(BL)\approx 0.2$, i.e., $L\sim 1000$, which is just slightly beyond the system sizes we can currently attain with reasonable accuracy.

\section{Surface correlation functions}
\label{sec:corr}

In this Section we investigate the surface correlation function $C_L(l)$. We start our analysis with the $S=1$ chains, ${\cal H}_{\rm AKLT}$ and ${\cal H}_1$, where the surface order and the localization of edge spins are well-understood. These models serve as benchmarks, and we ask to what extent these results can be generalized to the half-integer spin models, ${\cal H}_{3/2}$ and ${\cal H}_{1/2}^1(\lambda)$.

\subsubsection{$S=1$ spin chain}

Correlation functions can be calculated rigorously for the AKLT model, ${\cal H}_{\rm AKLT}$, where the exact ground state is the VBS state.\cite{AKLT,fath_jpc93,mikeska_csf95,polizzi_prb98} It turns out that the two point correlation function between any two points $i$ and $j>i$ in a finite open chain with length $L$ in the singlet ground state is
\begin{eqnarray}\label{AKLTcorr}
    &&\langle S_i^zS_j^z \rangle_{\rm AKLT} = \\&&\qquad
    \frac{4}{3\left[1-(-\frac{1}{3})^{L-1}\right]}
    \left[ \left(-\frac{1}{3} \right)^{j-i} + \left(-\frac{1}{3} \right)^{L-j+i} \right].\nonumber
\end{eqnarray}
From this the surface correlation function with $i=l$, $j=L+1-l$ reads
\begin{eqnarray}\label{AKLTcorr2}
    &&C_L(l) = \\&&\qquad
    \frac{4}{3\left[1-(-\frac{1}{3})^{L-1}\right]}
    \left[ \left(-\frac{1}{3} \right)^{L+1-2l} \!\!+ \left(-\frac{1}{3} \right)^{2l-1} \right].\nonumber
\end{eqnarray}
For $L\to\infty$ there is a finite surface order,
\begin{equation}\label{AKLTcorr3}
    C(l) = -4 \left(\frac{1}{3} \right)^{2l} = -4 e^{-2l/\xi_{\rm AKLT}}.
\end{equation}
with $\xi_{\rm AKLT}=1/\ln(3)$ the (bulk) correlation length of the AKLT model. Note that
$C(1)=\lim_{L\to\infty}C_L(1)= -4/9$, which is exactly equal in magnitude to the string order parameter. This is a peculiarity of the AKLT model, but there is a general qualitative relationship between the existence of end spins, the string order in the bulk, the surface order, and the non-vanishing entanglement between the end spins,\cite{verstraete_prl04} all holding for the pure $S=1$ chain too. All these quantities can be used as order parameters in the Haldane phase.

As expected, the $l$-dependence of $C_l$ and also the finite-size corrections are determined by the bulk correlation length $\xi_{\rm AKLT}$. For instance, for $C_L(1)$ we find the expansion
\begin{eqnarray}\label{AKLTcorr4}
    C_L(1) &=&
      -\frac{4}{9} - (-)^L \frac{8}{3} e^{-L/\xi_{\rm AKLT}} + \\[.1cm] &&
      +8\, e^{-2L/\xi_{\rm AKLT}} +{\cal O}(e^{-3 L/\xi_{\rm AKLT}}). \nonumber
\end{eqnarray}

For ${\cal H}_1$ we determined the surface order numerically. Precise numerical estimates for $C(l)$ and $\xi_f$ are obtained in the
following way. Assuming finite-size corrections in the form of Eq.\
(\ref{C_1_L}) we obtained effective, finite-size estimates from
three-point fit both for $C(l,L)$ and $\xi_f(L)$.


The fitted values for $C(l,L)$ converge very rapidly, and we find a clear surface order $C(l)=\lim_{L\to\infty} C(l,L)>0$, whose value for $l=1$ reads $C(1) = -0.28306484(1)$. Note for comparison that the string order parameter in this case is $0.374335$.\cite{white_huse}
The correlation length has a stronger $L$ dependence, which can be a result of a neglected algebraic prefactor in the r.h.s.\ of Eq.\ (\ref{C_1_L}). Its extrapolated value for the end-to-end correlation is $\xi_f=6.1(1)$, which should be equal to the bulk correlation length $\xi=6.03$,\cite{white_huse} as expected on general grounds.

\begin{figure}[tbp]
\centerline{\includegraphics[scale=.34]{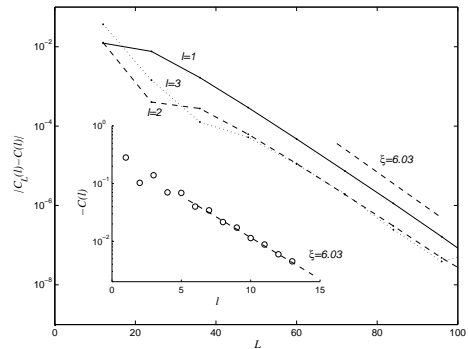}} 
\caption{Exponential finite-size scaling of the surface correlation function $C_L(l)$ of the $S=1$ HAF chain for $l=1,2,3$. Inset shows the behavior of the extrapolated values $C(l)$ as a function of $l$.}
\label{Fig2}
\end{figure}

The finite-size corrections to the surface spin correlations are
illustrated in Fig.\ \ref{Fig2}, in which $C_L(l)-C(l)$ for $l=1,2,3$
is plotted as a function of $L$ in a semi-logarithmic scale. Having an
appropriate extrapolation value $C(l)$, the points lie on an
approximately straight line, the asymptotic slope of which defines the
inverse of $\xi_f$. The slope of all
lines in Fig.\ \ref{Fig2} are compatible with  the bulk correlation length.

The extrapolated values of the surface correlation function $C(l)$,
obtained for different $l$-s are plotted in the inset of Fig.\
\ref{Fig2} in a log-log plot. The asymptotic slope is again compatible
with the known bulk correlation length. An accurate estimate is obtained by
calculating effective $\xi(l)$ values by two-point fits, see Table
\ref{Table3}, which is then extrapolated for $l\to\infty$ as $\xi_s=6.05(3)$.

\begin{table}
\caption{The surface correlation function, $C(l)$, of the $S=1$ chain and the (surface)
correlation length, $\xi_s$ obtained by two-point fit. The extrapolated values, $\xi^{ext}_s$,
are obtained by the BST convergence acceleration algorithm\cite{BS,carlon-igloi} with the parameter, $\omega=2.5$. \label{Table3}}
 \begin{tabular}{|c|c|c|c|}  \hline
$l$  & $C(l)$   & $\xi_s$     & $\xi^{ext}_s$   \\ \hline
1  & 0.28306484 &             &                 \\
3  & 0.13931623 & 5.64230917  &                 \\
5  & 0.06887108 & 5.67770491  & 5.69806489      \\
7  & 0.03439726 & 5.76153408  & 5.84348388      \\
9  & 0.01734266 & 5.84105901  & 5.95103010      \\
11 & 0.00880481 & 5.90082421  & 6.00726945      \\
13 & 0.00449158 & 5.94270505  & 6.03378854      \\  \hline

  \end{tabular}
  \end{table}

\subsubsection{$S=3/2$ spin chain}

The numerically determined values of the surface correlation function $C_L(l)$ for $l=1,2,3$ are given in Table \ref{Corr32_Table}. We believe that these values are accurate up to at least 8 digits.

\begin{table}
\caption{The surface correlation function $C_L(l)$ of the $S=3/2$ chain for $l=1,2,3$. \label{Corr32_Table}}
 \begin{tabular}{|c|c|c|c|}  \hline
$L$  & $C_L(1)$   & $C_L(2)$     & $C_L(3)$   \\ \hline
   12   &   -0.6149116033  &   -0.3463729630   &  -0.4577771615 \\
   24   &   -0.5312680335  &   -0.2796013080   &  -0.3483003563 \\
   36   &   -0.4960012887  &   -0.2563400653   &  -0.3154186768 \\
   48   &   -0.4754275850  &   -0.2437536782   &  -0.2984765312 \\
   60   &   -0.4614763573  &   -0.2355733053   &  -0.2877411795 \\
   72   &   -0.4511552246  &   -0.2296862928   &  -0.2801355963 \\
   84   &   -0.4430747706  &   -0.2251658062   &  -0.2743575096 \\
   96   &   -0.4364949818  &   -0.2215368743   &  -0.2697544487 \\
  108   &   -0.4309814145  &   -0.2185291971   &  -0.2659614093 \\
  120   &   -0.4262589357  &   -0.2159752350   &  -0.2627549925 \\
  132   &   -0.4221438229  &   -0.2137653001   &  -0.2599904263 \\
  144   &   -0.4185068985  &   -0.2118233164   &  -0.2575681425 \\
  156   &   -0.4152559777  &   -0.2100956569   &  -0.2554183252 \\
  168   &   -0.4123218910  &   -0.2085426652   &  -0.2534897645 \\
  180   &   -0.4096518527  &   -0.2071345489   &  -0.2517442428 \\
  192   &   -0.4072038977  &   -0.2058474800   &  -0.2501511731 \\ \hline
  \end{tabular}
  \end{table}

The fundamental question we would like to answer is whether $C(l)=\lim_{L\to\infty} C_L(l)$ remains finite as for the $S=1$ chain or scales to zero. Since bulk correlations are algebraic, we expect that neither $C_L(l)$ can contain a finite characteristic length, in clear contrast to the behavior of the
$S=1$ chain.

To obtain possible working hypotheses on $C_L(l)$, let us briefly consider an exactly soluble model, the \emph{Ising quantum chain with an extended defect},\cite{hilhorst_prl81} whose properties are summarized in the Appendix. For this model there are three different type of surface
critical behavior:

i) For weak enough defects $C(1)$ is zero and the end-to-end correlations vanish algebraically. In this case the scaled gap, $L \Delta_1$, has a finite limiting value predicted by conformal invariance.

ii) For a critical strength of the defect $C(l)$ is zero and the end-to-end correlations decay logarithmically. In this case the scaled gap vanishes logarithmically.

iii) Finally, for strong enough defects $C(l)$ is finite, which is approached by algebraic corrections. At the same time the scaled first gap goes to zero algebraically with $L$.

\begin{figure}[tbp]
\centerline{\includegraphics[scale=.5]{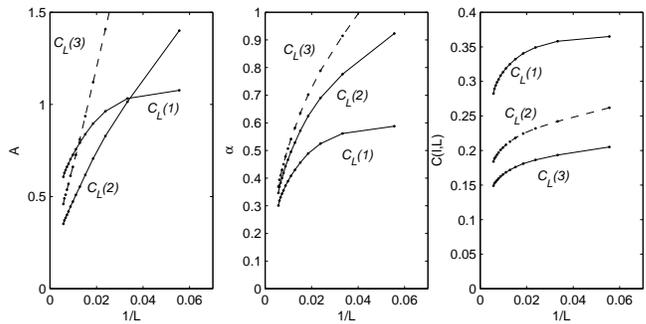}} 
\caption{Local fit parameters for Eq.\ (\ref{badfit}) calculated from three-point fits.}
\label{Fig_badfit}
\end{figure}

We start to analyse surface correlations for the $S=3/2$ spin chain using an algebraic
ansatz
\begin{equation}\label{badfit}
    C_L(l) = C(l) + A L^{-\alpha}\;,
\end{equation}
which works for regions i) and iii) of the inhomogeneous Ising quantum chain.
Local approximations to the three fit parameters can be determined from three-point fits using the consecutive numerical values for $C_{L-12},C_{L}$ and $C_{L+12}$. These $L$-dependent local parameters are depicted in Fig.\ \ref{Fig_badfit} as a function of $1/L$. If the ansatz were good, the fit parameters should stabilize as $L$ increases. However, as is clearly seen in the Figures, all parameters are monotonically decreasing functions of $L$, and there is no sign of convergence. The effective exponent $\alpha(L)$ reaches 0.3 for $L=192$, and it continues to fall rapidly. The effective values of $\alpha$ are much smaller than the known surface exponent of the $S=1/2$ chain. In fact, the rapid decrease seems to be consistent with a vanishing exponent in the $L\to\infty$ limit. Approximants for $C(1)$ also decrease considerably for increasing $L$. The most probable interpretation is that the ansatz in Eq.\ (\ref{badfit}) is inadequate. Our experience was similar with $C_L(2)$ and $C_L(3)$ as is shown in Fig.\ \ref{Fig_badfit}.

\begin{figure}[tbp]
\centerline{\includegraphics[scale=.45]{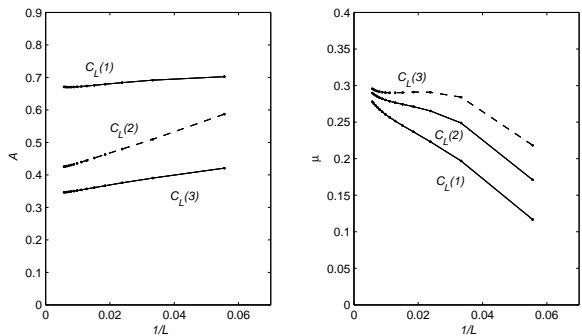}} 
\caption{Local fit parameters for Eq.\ (\ref{goodfit}) calculated from two-point fits for the ${\cal H}_{3/2}$ model.}
\label{Fig_goodfit}
\end{figure}

Now we try to analyze the data instead of Eq.\ (\ref{badfit}) with a purely logarithmic behavior with vanishing $C(l)$,
\begin{equation}\label{goodfit}
    C_L(l) = -A_l \ln(BL)^{-\mu},
\end{equation}
which contains two free parameters $A_l$ and $\mu$, if we use the value of $B$ determined from the first gap. This form corresponds to the borderline case ii) for the inhomogeneous Ising quantum chain. This type of analysis is supported by the fact that the scaled first gap is vanishing logarithmically in both models. Indeed, this fit describes the correlation data more satisfactorily, as is obvious from the stability of the local parameters in Fig.\ \ref{Fig_goodfit}.

In order to present a theoretical basis to this procedure we note that
having a  marginal surface operator which scales as in Eq.\ (\ref{g_scales}), the end-to-end Green's function, $G^{(n)}_L=\langle \phi_n(0) \phi_n(L) \rangle$, associated with operator $\phi_n$, is believed to satisfy the Callan-Symanzik equation\cite{AGST,Nomura}
\begin{equation}\label{}
    \left[L\frac{\partial}{\partial L}+\beta(g) \frac{\partial}{\partial g}
    +2 \gamma_n(g)  \right]G^{(n)}_L =0,
\end{equation}
where the beta and gamma functions are
\begin{equation}\label{}
    \beta(g) = -\pi b g^2;\quad  \gamma_n(g) = x_n + 2\pi b_n g.
\end{equation}
From this, up to ${\cal O}(g)$, the Greens function reads\cite{AGST}
\begin{equation}\label{callan-syma}
    G^{(n)}_L = \textrm{const}\; L^{-2x_n} \left[ \frac{1}{\pi b g(L)}  \right]^{-4b_n/b}.
\end{equation}
We have seen that for the first excited (boundary) state $x_1=0$. Using then the actual scaling form of $g(L)$ in Eq.\ (\ref{g_scales}), we obtain
\begin{equation}\label{}
    G^{(1)}_L \sim \ln(BL)^{-2 d_1},
\end{equation}
where we have used that by Eq.\ (\ref{dn}) $d_1=2b_1/b$.\cite{AGST} Expecting that for small $l$ the leading contribution to $C_L(l)$ is given by $G^{(1)}_L$ we obtain the fitting ansatz in Eq.\ (\ref{goodfit}).

The results for $A_l$ and $\mu$ are shown in Fig.\ \ref{Fig_goodfit}. For $l=1,2,3$ the measured exponent $\mu$ is estimated to be $\mu=0.31\pm 0.02$. This value is convincingly close to $2d_1=0.28\pm 0.04$, obtained earlier from the gap, supporting the theory outlined above. We note that the exact results of the inhomogeneous Ising quantum chain in case ii) are in complete agreement with the predicted behavior. The prefactor of the logarithmically vanishing scaled gap in Eq.\ (\ref{Isinggap}) is just the half of the exponent of the logarithmic decay of the end-to-end correlations in Eq.\ (\ref{IsingC}).

\subsubsection{$S=1/2$ spin chain with impurity end spins}

\begin{figure}[tbp]
\centerline{\includegraphics[scale=.45]{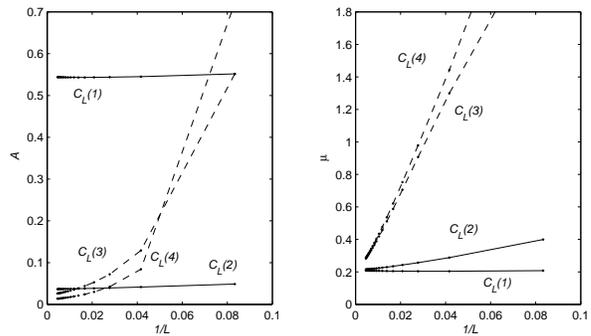}} 
\caption{Local fit parameters for Eq.\ (\ref{goodfit}) calculated from two-point fits for the ${\cal H}_{1/2}^1(\lambda=1)$ model.}
\label{Fig_goodfit_imp}
\end{figure}

Finally we analyze the surface correlations for the impurity model ${\cal H}_{1/2}^1(\lambda)$ with $\lambda=1$.  The fit parameters for Eq.\ (\ref{goodfit}) are presented in Fig.\ \ref{Fig_goodfit_imp}. Scaling to finite values is very convincingly satisfied, especially for the end-to-end correlation, $C_L(1)$. There the extrapolated exponent is $\mu=0.21$ and the amplitude is $A=0.54$. The exponent is found to be the same with somewhat higher error bars for any $l$ in $C_L(l)$, and the amplitude $A_l$ is a monotonically decreasing function of $l$. Again the exponent nicely satisfies the expected relation $\mu=2d_1$ with $d_1=0.12\pm 0.02$ from the first gap.

\section{Discussion}
\label{sec:disc}

In this paper we have studied the energy spectrum and the surface spin correlations of the $S=3/2$ antiferromagnetic Heisenberg chain. The analysis of the numerical data for both quantities was
difficult due to strong logarithmic terms, which are the consequence of the
presence of a bulk and a surface marginal operator in the fixed point. Our very
accurate numerical data calculated by an advanced DMRG algorithm on finite open chains
made it possible to infer the true asymptotic behavior. The end-to-end correlations
are found to approach a vanishing limiting value, however the decay is extremely
slow, it is logarithmic in $L$. This type of logarithmic decay is accompanied by
a logarithmically vanishing scaled first gap in the spectrum of the system. This
latter result is in accordance with a field-theoretical analysis of the problem,
which is based on the assumption of localized edge states as in the $S=1$ chain. Localization of
edge spins, however, would imply an exponentially decaying surface correlation function,
which is in contradiction with our numerical results. We can thus say that the edge
states are logarithmically delocalizing, or framing it differently, are just ``quasi-localized".
Having in mind the analogous result for the
inhomogeneous Ising quantum chain in Eq.\ (\ref{Isingpsi}) we can say that the edge spins are
concentrated in a region of size $\sim \ln L$ at the end of the chain. The delocalization of edge spins is so weak that even for a macroscopic system we may have the illusion that they exist and can be measured experimentally. For instance, we have seen that for $L=108$ the end-to-end correlation is $C_L(1)\approx -0.43$, while for a macroscopic system with $L=10^{23}$ spins the fitted scaling form implies $C_L(1)\approx -0.18$, still a relatively large value.

For comparison we have also analyzed the gaps and surface correlations of an $S=1/2$ chain with $S=1$ impurity end spins, which was predicted to be equivalent in the VBS picture. Indeed, we found that the exponent of the logarithmic decay of the correlations and the prefactor of the logarithmically vanishing scaled first gap are universal.

Higher gaps of the spectrum, in particular the second gap, scale as $1/L$.
The precise analysis of the second gap is numerically more demanding. Whereas
we were able to check and more or less confirm some existing conjectures for the
impurity models, we could not convincingly reach the thermodynamic
limit for the $S=3/2$ chain. Up to the lengths we could achieve in our
DMRG calculation the anomalous dimension $x_2$ shows a non-trivial
value strictly below the predicted exponent $x_2=1$. More work is needed in the
future on this issue.

We close our paper with some remarks.
First, we note that for not too long chains the numerical data
can be easily misinterpreted to have an effective gap exponent, $\zeta>1$,
and a finite limiting end-to-end correlation function, which is approached
as a power-law. [For the analogous inhomogeneous Ising quantum chain this
happens for strong defect couplings, $A>1$, see in Eqs.\ (\ref{Isinggap})
and (\ref{IsingC}).] This type of behavior was suggested in an earlier numerical
work,\cite{CLRI} but data on longer chains gives evidence for the
logarithmic dependence.

Half-integer spin chains with $S > 3/2$ are expected to have the same type
of logarithmic end-to-end correlations as the $S=3/2$ chain. In these cases
the exponent of the logarithmic decay, $2d$, is probably $S$ dependent.
We wonder if a field-theoretical approach, such as logarithmic conformal
field theory can predict the values of these exponents.

Our final remark concerns the effect of quenched disorder. The
quasi surface order in the $S=3/2$ chain is found to be robust against weak
quenched disorder. As studied in Ref.\ \onlinecite{CLRI} the end-to-end
correlations show the same qualitative behavior for not too strong randomness,
and at the same
time the critical properties of the system remains the same as in the
non-random chain. For stronger, but still finite disorder the (quasi) surface
order vanishes, and the critical behavior of the system is
controlled by new types of disorder fixed points.\cite{CLRI,fisher}

\begin{acknowledgments}
The authors are grateful to Z. Bajnok, E.\ Carlon, 
J.\ S\'olyom, and L.\ Turban for useful discussions. This work has been supported by a German-Hungarian exchange program (DAAD-M\"OB), by the
Hungarian National Research Fund under grant Nos. OTKA TO34138, TO37323, T04330, F046356,
TO48721, MO45596 and M36803, and by Kuwait University Research Grant No.\ SP 09/02. The authors acknowledge computational support by Dynaflex
Ltd under Grant No.\ IgB-32 and by NIIF under Grant Nos. 1030 and 1038. \"O.\ L.\ was also supported by the J\'anos Bolyai research scholarship.
\end{acknowledgments}

\appendix

\section{Surface critical behavior of the Ising quantum chain with an extended defect}
\label{app:HvL}

Here we consider an inhomogeneous Ising quantum chain, for which many properties
such as the end-to-end correlation function and the excitation spectrum can be
calculated exactly,\cite{hilhorst_prl81,IPT} and these show similar characteristics as those of the $S=3/2$ chain. These exact results are expected to serve as an analogy in the analysis.

The inhomogeneous quantum Ising chain is defined by the Hamiltonian
\be
H_I=-\frac{1}{2}\left[ \sum_{i=1}^{L-1} J_l \sigma_l^z \sigma_{l+1}^z + \sum_{i=1}^{L}  \sigma_l^x \right]\;,
\label{Ising}
\ee
with a smoothly varying coupling
\be
J_l=1+\frac{A}{2} \left[ \frac{L}{\pi} \sin \left(\frac{\pi l}{L} \right) \right]^{-1}\;,
\label{HvL}
\ee
in terms of the Pauli-spin operators, $\sigma_l^{x,z}$, at site $l$.
This quantum Hamiltonian is related through a conformal transformation
to a classical two-dimensional semi-infinite Ising model, the couplings of
which at a distance $l$ from the free surface deviate from the bulk value by an
amount of $A/2l$.

Here we list exact results about the end-to-end correlation function,
$C_L(1)=\langle \sigma_1^x \sigma_L^x \rangle$, as well as for the lowest gap $\epsilon_1$,
which can be calculated in the free-fermion representation, when the Hamiltonian is expressed
as\cite{IPT}
\be
H_{I}=\sum_{q} \epsilon_q (\eta_q^{+} \eta_q-1/2)\;,
\label{H_FF}
\ee
in terms of fermion creation, $\eta_q^{+}$, and annihilation, $\eta_q$, operators. The
energy of the fermionic excitation, $\epsilon_q$, is obtained from the solution of an
$L \times L$ eigenvalue equation: $(\bf{A}+\bf{B})(\bf{A}-\bf{B})\psi_q=\epsilon_q^2 \psi_q$,
where the eigenvector, $\psi_q$, contains information about the localized or delocalized nature of
the excitation.

The quantum Ising model in Eq.\ (\ref{Ising}) has the same type of bulk quasi-long-range order
as the homogeneous chain with $A=0$, but the surface properties of the system are
modified by the smoothly varying inhomogeneity. These properties are qualitatively different
for $A <1$, $A=1$ and for $A > 1$.

The behavior of the end-to-end correlation function can be summarized as follows:
\be \label{IsingC}
C_L(1)=\left\{
\begin{array}{ll}
  {\rm const}\, L^{A-1} &\mbox{if $A<1$} \\
  {\rm const} \ln^{-1}(L) & \mbox{if $A=1$} \\
  m_s^2 + {\rm const}\, L^{1-A}  & \mbox{if $A>1$}, \\
\end{array}
\right.
\ee
i.e., it tends to zero as a power-law for $A<1$, logarithmically for $A=1$, and to
a finite value with power-law corrections for $A>1$.

The first gap scales as
\be \label{Isinggap}
\epsilon_1=\left\{
\begin{array}{ll}
  \dfrac{\pi}{L} \dfrac{1-A}{2} &\mbox{if $A<1$} \\[.3cm]
  \dfrac{1}{2}\dfrac{\pi}{L} \dfrac{1}{\ln L} & \mbox{if $A=1$} \\[.3cm]
  {\rm const}\, L^{-A}  & \mbox{if $A>1$}. \\
\end{array}
\right.
\ee
The gap vs exponent relation established by Eqs.\ (\ref{cft}), (\ref{dn}) and (\ref{callan-syma}) is satisfied for $A<1$ and $A=1$. For $A>1$ the first gap is determined by a mode localized to the surface. This latter becomes clear when the eigenvector of the lowest excitation, $(-)^l\psi_1(l)$ is considered, which is given in the continuum approximation in terms of $z=\pi l/L$ as
\be \label{Isingpsi}
\psi_1(z)=\left\{
\begin{array}{ll}
  \dfrac{1}{L^{1/2}} \dfrac{\sin^{1-A/2}(z/2)}{\cos^{A/2}(z/2)} &\mbox{if $A<1$} \\[.4cm]
  \dfrac{\ln(zL)}{L^{1/2}(\ln L)^{3/2}} \tan^{1/2}(z/2) & \mbox{if $A=1$} \\[.4cm]
  \dfrac{1}{L^{A/2}}\tan^{A/2}(z/2)  & \mbox{if $A>1$}. \\
\end{array}
\right.
\ee
This eigenvector is localized at $l=L$ for $A>1$, logarithmically delocalized (quasi-localized) for $A=1$ and delocalized for $A<1$. Note that another surface excitation can be constructed, which is localized at $l=1$, as dictated by symmetry.

\end{document}